\documentclass[
 preprint,
 superscriptaddress,
 amsmath,amssymb,
 aps,
 pra,
 longbibliography,
]{revtex4-1}

\usepackage{graphicx}% Include figure files
\usepackage{dcolumn}% Align table columns on decimal point
\usepackage{bm}% bold math
\usepackage [latin1]{inputenc}

\begin{document}

\title{Strong light-matter interactions between gap plasmons and two-dimensional excitons at ambient condition in a deterministic way}

\author{Longlong Yang}
\affiliation{Beijing National Laboratory for Condensed Matter Physics, Institute of Physics, Chinese Academy of Sciences, Beijing 100190, China}
\affiliation{CAS Center for Excellence in Topological Quantum Computation and School of Physical Sciences, University of Chinese Academy of Sciences, Beijing 100049, China}
\author{Xin Xie}
\affiliation{Beijing National Laboratory for Condensed Matter Physics, Institute of Physics, Chinese Academy of Sciences, Beijing 100190, China}
\affiliation{CAS Center for Excellence in Topological Quantum Computation and School of Physical Sciences, University of Chinese Academy of Sciences, Beijing 100049, China}
\author{Jingnan Yang}
\affiliation{Beijing National Laboratory for Condensed Matter Physics, Institute of Physics, Chinese Academy of Sciences, Beijing 100190, China}
\affiliation{CAS Center for Excellence in Topological Quantum Computation and School of Physical Sciences, University of Chinese Academy of Sciences, Beijing 100049, China}
\author{Mengfei Xue}
\affiliation{Beijing National Laboratory for Condensed Matter Physics, Institute of Physics, Chinese Academy of Sciences, Beijing 100190, China}
\affiliation{CAS Center for Excellence in Topological Quantum Computation and School of Physical Sciences, University of Chinese Academy of Sciences, Beijing 100049, China}
\author{Shiyao Wu}
\affiliation{Beijing National Laboratory for Condensed Matter Physics, Institute of Physics, Chinese Academy of Sciences, Beijing 100190, China}
\affiliation{CAS Center for Excellence in Topological Quantum Computation and School of Physical Sciences, University of Chinese Academy of Sciences, Beijing 100049, China}
\author{Shan Xiao}
\affiliation{Beijing National Laboratory for Condensed Matter Physics, Institute of Physics, Chinese Academy of Sciences, Beijing 100190, China}
\affiliation{CAS Center for Excellence in Topological Quantum Computation and School of Physical Sciences, University of Chinese Academy of Sciences, Beijing 100049, China}
\author{Feilong Song}
\affiliation{Beijing National Laboratory for Condensed Matter Physics, Institute of Physics, Chinese Academy of Sciences, Beijing 100190, China}
\affiliation{CAS Center for Excellence in Topological Quantum Computation and School of Physical Sciences, University of Chinese Academy of Sciences, Beijing 100049, China}
\author{Jianchen Dang}
\affiliation{Beijing National Laboratory for Condensed Matter Physics, Institute of Physics, Chinese Academy of Sciences, Beijing 100190, China}
\affiliation{CAS Center for Excellence in Topological Quantum Computation and School of Physical Sciences, University of Chinese Academy of Sciences, Beijing 100049, China}
\author{Sibai Sun}
\affiliation{Beijing National Laboratory for Condensed Matter Physics, Institute of Physics, Chinese Academy of Sciences, Beijing 100190, China}
\affiliation{CAS Center for Excellence in Topological Quantum Computation and School of Physical Sciences, University of Chinese Academy of Sciences, Beijing 100049, China}
\author{Zhanchun Zuo}
\affiliation{Beijing National Laboratory for Condensed Matter Physics, Institute of Physics, Chinese Academy of Sciences, Beijing 100190, China}
\affiliation{CAS Center for Excellence in Topological Quantum Computation and School of Physical Sciences, University of Chinese Academy of Sciences, Beijing 100049, China}
\author{Jianing Chen}
\email{jnchen@iphy.ac.cn}
\affiliation{Beijing National Laboratory for Condensed Matter Physics, Institute of Physics, Chinese Academy of Sciences, Beijing 100190, China}
\affiliation{CAS Center for Excellence in Topological Quantum Computation and School of Physical Sciences, University of Chinese Academy of Sciences, Beijing 100049, China}
\affiliation{Songshan Lake Materials Laboratory, Dongguan, Guangdong 523808, China}

\author{Yuan Huang}
\email{yhuang01@iphy.ac.cn}
\affiliation{Beijing National Laboratory for Condensed Matter Physics, Institute of Physics, Chinese Academy of Sciences, Beijing 100190, China}
\affiliation{CAS Center for Excellence in Topological Quantum Computation and School of Physical Sciences, University of Chinese Academy of Sciences, Beijing 100049, China}
\affiliation{Songshan Lake Materials Laboratory, Dongguan, Guangdong 523808, China}
\author{Xingjiang Zhou}
\affiliation{Beijing National Laboratory for Condensed Matter Physics, Institute of Physics, Chinese Academy of Sciences, Beijing 100190, China}
\affiliation{CAS Center for Excellence in Topological Quantum Computation and School of Physical Sciences, University of Chinese Academy of Sciences, Beijing 100049, China}
\affiliation{Songshan Lake Materials Laboratory, Dongguan, Guangdong 523808, China}

\author{Kuijuan Jin}
\affiliation{Beijing National Laboratory for Condensed Matter Physics, Institute of Physics, Chinese Academy of Sciences, Beijing 100190, China}
\affiliation{CAS Center for Excellence in Topological Quantum Computation and School of Physical Sciences, University of Chinese Academy of Sciences, Beijing 100049, China}
\affiliation{Songshan Lake Materials Laboratory, Dongguan, Guangdong 523808, China}

\author{Can Wang}
\email{canwang@iphy.ac.cn}
\affiliation{Beijing National Laboratory for Condensed Matter Physics, Institute of Physics, Chinese Academy of Sciences, Beijing 100190, China}
\affiliation{CAS Center for Excellence in Topological Quantum Computation and School of Physical Sciences, University of Chinese Academy of Sciences, Beijing 100049, China}
\affiliation{Songshan Lake Materials Laboratory, Dongguan, Guangdong 523808, China}

\author{Xiulai Xu}
\email{xlxu@iphy.ac.cn}
\affiliation{Beijing National Laboratory for Condensed Matter Physics, Institute of Physics, Chinese Academy of Sciences, Beijing 100190, China}
\affiliation{CAS Center for Excellence in Topological Quantum Computation and School of Physical Sciences, University of Chinese Academy of Sciences, Beijing 100049, China}
\affiliation{Songshan Lake Materials Laboratory, Dongguan, Guangdong 523808, China}

%%%%%%%%%%%%%%%%%%%%%%%%%%%%%%%%%%%%%%%%%%%%%%%%%%%%%%%%%%%%%%%%%%%%%
%% Some journals require a list of abbreviations or keywords to be
%% supplied. These should be set up here, and will be printed after
%% the title and author information, if needed.
%%%%%%%%%%%%%%%%%%%%%%%%%%%%%%%%%%%%%%%%%%%%%%%%%%%%%%%%%%%%%%%%%%%%%

%\keywords{strong coupling, transition metal dichalcogenides, exciton, gap plasmon, effective exciton number   \LaTeX}

\date{\today}

\begin{abstract}
Strong exciton-plasmon interaction between the layered two-dimensional (2D) semiconductors and gap plasmons shows a great potential to implement cavity quantum-electrodynamics in ambient condition. However, achieving a robust plasmon-exciton coupling with nanocavity is still very challenging, because the layer area is usually small with conventional approaches. Here, we report on a robust strong exciton-plasmon coupling between the gap mode of bowtie and the excitons in MoS$_2$ layers with gold-assisted mechanical exfoliation and the nondestructive wet transfer techniques for large-area layer. Benefiting from the ultrasmall mode volume and strong in-plane field, the estimated effective exciton number contributing to the coupling is largely reduced. With a corrected exciton transition dipole moment, the exciton numbers are extracted with 40 for the case of monolayer and 48 for 8 layers. Our work paves a way to realize the strong coupling with 2D materials with few excitons at room temperature.
\end{abstract}
%$\bf{keywords}$: strong coupling, transition metal dichalcogenides, exciton, gap plasmon, effective exciton number

\maketitle
\section{Introduction}

 Atomically thin transition metal dichalcogenides (TMDs) have been exploited widely for numerous optoelectronics and photonics applications, including single-photon emitters \cite{chakraborty2015voltage,dang2020identifying}, transistors \cite{zhang2012ambipolar,radisavljevic2011single}, photodetectors \cite{mak2014valley,lopez2013ultrasensitive} and valleytronic devices \cite{xiao2012coupled,cao2012valley}. One of the intriguing properties is the large exciton binding energy \cite{ugeda2014giant,he2014tightly}, providing the opportunity to study the interaction of light and matter at room temperature when they are embedded in a microcavity \cite{lundt2016room,wang2016giant}. When the rate of coherent energy transfer between excitonic transition and photons in cavity is faster than their average dissipation rate, the system goes into strong coupling regime, resulting in the formation of part-light and part-matter bosonic quasiparticles called microcavity polaritons \cite{liu2015strong,torma2014strong}. Polaritons in microcavities provide excellent platforms to realize Bose-Einstein condensation \cite{deng2002condensation,kasprzak2006bose}, low-threshold lasing \cite{christopoulos2007room,kena2010room}, low-energy switching \cite{dreismann2016sub,ballarini2013all} and quantum information processing \cite{wallraff2004strong,amo2010exciton}.

 %large exciton binding energy\cite{ugeda2014giant,mak2012control,he2014tightly},
 %embedded in the microcavity\cite{lundt2016room,wang2016giant,sun2017optical,li2017room}.

 %condensation,\cite{deng2002condensation,kasprzak2006bose,deng2010exciton,waldherr2018observation} low-threshold
 %\cite{chakraborty2015voltage,koperski2015single,srivastava2015optically,luo2018deterministic,he2015single,dang2020identifying}
 %optical nonlinear process \cite{khitrova1999nonlinear,sukharev2018effects}

In order to achieve strong coupling with excitons in TMDs, optical cavities such as Fabry$-$P$\acute{e}$rot cavities and photonic crystal microcavities \cite{xie2020topological,qian2019} have been employed widely at cryogenic temperatures and in high vacuum \cite{dufferwiel2015exciton,chen2017valley}. Even though a few of them have been used to demonstrate strong coupling at room temperature \cite{liu2015strong,zhang2018photonic,flatten2016room}, the Rabi splittings are on the order of thermal energy $k_{B}T$ ($\approx26$ meV) \cite{kleemann2017strong,hu2020recent}. Plasmonic nanocavities such as individual metallic nanoparticles \cite{chikkaraddy2016single} or dimers \cite{santhosh2016vacuum} can provide surface plasmon polaritons (SPPs) with light confined at subwavelength scale, which is an alternative choice for realizing strong coupling at ambient conditions, for example, coupled system with nanoparticle and layered TMDs \cite{kleemann2017strong,zheng2017manipulating,stuhrenberg2018strong,qin2020revealing,nanocube}.

The plasmonic nanocavities with mode volumes beyond the diffraction limit make it possible to demonstrate strong coupling with a small number of excitons, which has rich applications in the research of quantum many-body phenomena\cite{hartmann2008quantum}, photon blockade with many emitters\cite{trivedi2019photon}, cavity cooling \cite{hosseini2017cavity} and so on. Recently, the gap plasmon systems with ultimate field confinement have been used to realize strong coupling with single molecules \cite{chikkaraddy2016single} and single quantum dots \cite{santhosh2016vacuum}, indicating a potential for applications at the quantum optics level \cite{hennessy2007quantum,faraon2008coherent}. For layered TMDs, particle-on-film systems with nanoparticles \cite{Qi2020}, or single plasmonic structures such as nanorods \cite{SinglePlasmonicNanorod}, nanodisks \cite{Geisler2019} or nanocuboid \cite{Lo2019} have been used to demonstrate strong coupling. However, drop-casted nanoparitcles induce the randomness intrinsically, which cannot guarantee the robustness and the reproducibility in a deterministic way. The robustness is very important for the strong coupled system with a small number of excitons in particular \cite{liu2017strong,tame2013quantum}. In addition, the resonant optical field in such cavities is typically polarised perpendicularly to the layer planes when embedded with two-dimensional semiconductors \cite{kleemann2017strong,qin2020revealing}, which impedes the efficient coupling with the exciton dipole oriented almost in-plane. Therefore, engineering orientations of exciton dipole and cavity modes precisely is highly desired to large coupling strength with a small number of excitons.

To demonstrate the strong coupling in plasmon-exciton ("plexciton") with few excitons, the exciton number estimation is crucial. So far, the exciton number extraction with different ways shows a very large discrepancy \cite{SinglePlasmonicNanorod,qin2020revealing,nanocube,wang2019limits,zheng2017manipulating,liu2021plasmon}. Experimentally, the exciton number evolved in the strongly coupled system is related to the coupling strength, mode volume and the exciton transition dipole moment of TMDs \cite{qin2020revealing}. The coupling strength can be measured and the mode volume can be calculated with a good accuracy. It is noticed that the exciton transition dipole moment brings the main difference in previous work, which has been calculated with traditional quantum well recombination model \cite{asada1984gain} or extracted with absorption spectra \cite{zheng2017manipulating}. Here, we verify the two methods and correct the transition dipole moment values for the exciton number estimation.

In this work, we report on an observation of robust strong plasmon-exciton coupling between gap plasmons confined within individual gold bowties and excitons in MoS$_2$ layers at ambient conditions utilizing the gold-assisted mechanical exfoliation method and wet transfer techniques. Due to the strong in-plane electric field inside the material and small mode volume introduced by the bowtie resonator, vacuum Rabi splittings up to 110 meV and 80 meV are obtained for the coupling systems with 8-layer and monolayer MoS$_2$, respectively. The estimated effective exciton number $N$ contributing to the coupling with gap mode is reduced to $N \sim40$ for the case of monolayer and $N\sim48$ for the case of 8 layers with a corrected dipole moment. The robust strong plasmon-exciton coupling with less exciton number paves a way for future scalable integrated nanophotonic devices.

\section{Results and discussion}
\begin{figure}
\centering
\includegraphics[scale=0.5]{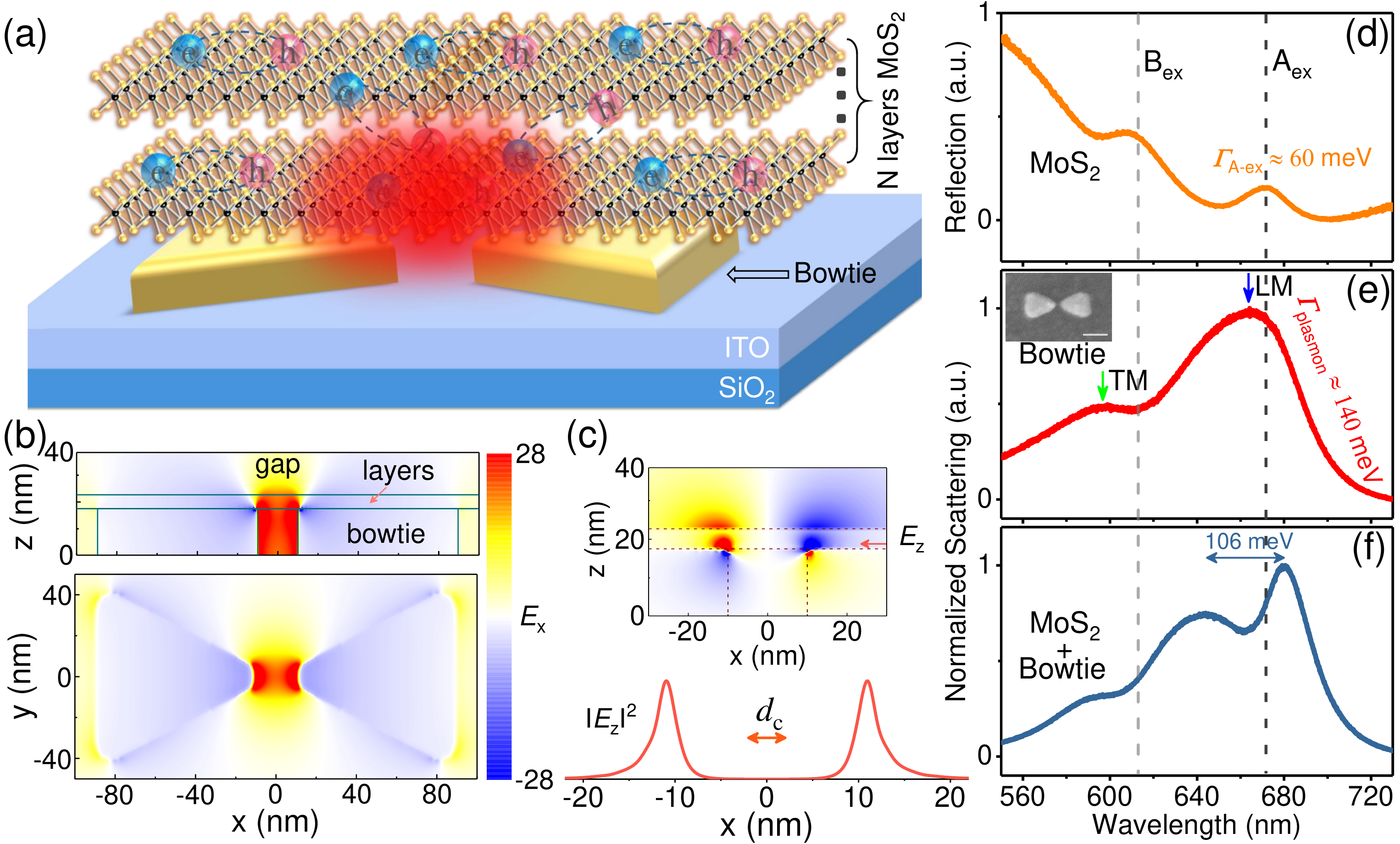} %scale is the size of the figure, you can modify it. f1_v1 is the name of figure.eps
\caption{TMDs excitons, gap plasmons, and strong coupling between them. (a) Schematic of the system with layered MoS$_2$ on a single gold bowtie resonator. (b)  X component of electric fields viewed from x-z (crosses along the middle line of bowtie) and x-y (within the layers) planes corresponding to gap resonances. (c) Upper: z component of electric fields viewed from x-z plane. Lower: distribution of electric field intensity along the position indicated by the magenta arrow. The baseline of the lower curve is zero. (d) Reflection spectrum of 8-layer MoS$_2$ on ITO/glass substrate, showing an A exciton peak with a line width of 60 meV. (e) Scattering spectrum of bowtie resonator, showing a well isolated longitudinal mode (LM) with line width of about 140 meV and transverse mode (TM). Inset shows a corresponding scanning electronic microscopy (SEM) image of bowtie nanostructure with a gap distance of  $\sim$ 20 nm (scale bar: 100 nm). (f) Scattering spectrum of a coupled hybrid system.}% the caption of figure
\label{p1}
%\label{f1}%the citation of the figure in the manuscript
\end{figure}

%\section{\label{sec2}Experiment and Discussion:\protect\\}
Figure \ref{p1}a shows a schematic diagram of the plasmon-exciton coupling system with layered MoS$_2$ on top of a bowtie nanostructure. Here, the gold bowtie is employed as a plasmonic nancavity for two reasons. First, it provides an ultra-confined gap plasmon mode with a mode volume of around $10^{3}$ $nm^3$ \cite{santhosh2016vacuum,lee2015fano}. More importantly, when the high-refractive-index MoS$_2$ layers is covered to the surface of bowtie, the strongly confined in-plane electric component of gap mode is expanding in the material (as shown Fig. \ref{p1}b) calculated with the finite-difference time-domain method, indicating the excitons above the gap will be strongly coupled to the gap mode. Normally in particle-on-film systems, the main electric field of the gap plasmon is dominated by out-of-plane component ($E_z$) \cite{kleemann2017strong,qin2020revealing} and strong coupling is achieved with the contribution of a large number of excitons. Our configuration is more suitable to enhance the coupling strength for the monolayer TMDs with completely in-plane dipole orientation or few layers with in-plane dipole strength dominated \cite{schuller2013orientation}.

To further compress the exciton number contributing to plasmon-exciton interaction, a localized electric field region comparable to the effective exciton area is required. The excitons in TMDs are delocalized quasiparticles formed in semiconductor band gaps and extending much larger than single unit cell \cite{baranov2018novel} with the exciton coherence length $d_c$ is $\sim4 $ nm for monolayer MoS$_2$ (Sec. III of the Supporting Information (SI)). Figure \ref{p1}c shows the distribution of out-of-plane component $E_z$ of the gap mode. It can be seen that the electric fields between two tips have opposite phases, but the intensities rapidly decay to zero inside the gap. Therefore, a gap distance slightly larger than the exciton coherence length should be able to couple the gap mode to few excitons with a constructive interference. In our device, the bowtie nanocavity was designed with a gap distance of about 20 nm. The corresponding dark-field (DF) scattering spectrum (Fig.~\ref{p1}e) shows a well-defined longitudinal gap mode at about 1.87 eV and a transverse mode at about 2.07 eV (Fig. S7 in the SI). Clearly, the longitudinal gap mode overlaps with the emission peak of A exciton of MoS$_2$ (Fig.~\ref{p1}d), satisfying the requirement of spectral coincidence.

Furthermore, a stable and reproducible coupled system is important to achieve strong coupling especially at few-exciton level, which has been an issue for systems based on colloidal quantum dots, molecules and TMDs \cite{chikkaraddy2016single,santhosh2016vacuum,hartsfield2015single,Yan2020} because of the randomness in exciton materials or plasmon nanocavities \cite{baranov2018novel}. Here, we have made two efforts to address this problem. First, a contamination-free, one-step and universal gold-assisted mechanical exfoliation method \cite{huang2020universal} has been used to obtain millimeter-size mono-/multilayer MoS$_2$ (Sec. I of the SI for more details), based on the covalent-like quasi bonding between Au adhesion layer and layered crystal. The quality of obtained MoS$_2$ layers is similar to that of the flakes prepared by traditional mechanical exfoliation, with clear Raman signals and photoluminescence from excitons (Fig. S4 and Fig. S5 in the SI). This exfoliation method with high yield ratio and large-area layers is essential for the rest systematic studies. Second, to guarantee the stability of the coupled systems, a nondestructive wetting transfer method has been used to transfer thin layers to the prepared nanocavities, without damaging the fragile nanostructures during the transfer process (see Sec. I of the SI for more details). As a result, a splitting of about 106 meV is achieved in a coupled system with resonators combined with 8-layer MoS$_2$ (as shown in Fig.~\ref{p1}f).

\begin{figure}
\centering
\includegraphics[scale=0.5]{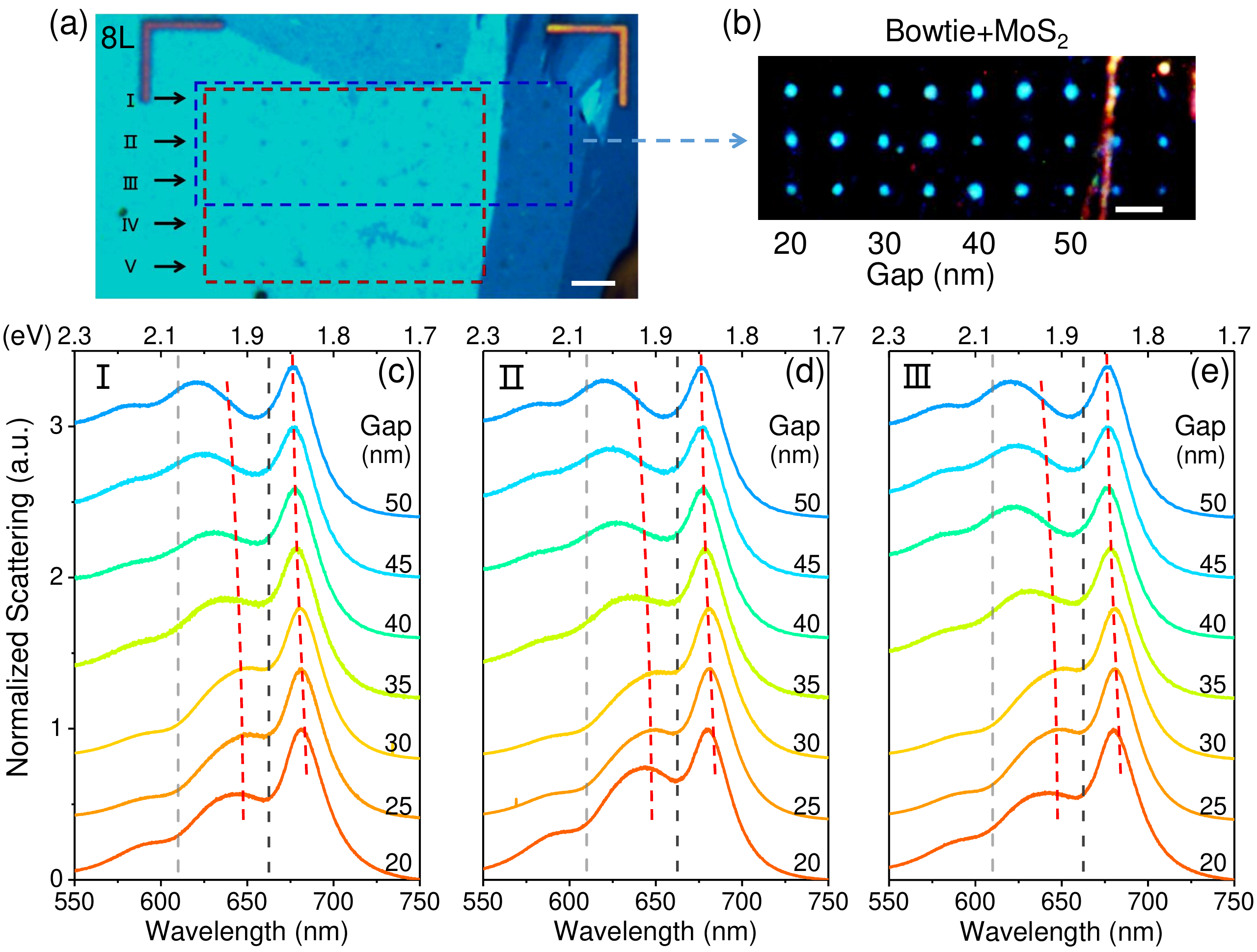} %scale is the size of the figure, you can modify it. f1_v1 is the name of figure.eps
\caption{\label{p2} Strong coupling of individual bowtie resonators with 8-layer MoS$_2$. (a) Bright-field image of MoS$_2$ layers on bowties (scale bar: 4 ${\mu}m$). The gap distance of bowtie increases from left to right and five rows of bowties with the same parameters are coated by MoS$_2$ layers (red dashed box). (b) Dark-field image of the first three rows in (a) (blue dashed box) (scale bar: 4 ${\mu}m$). (c) - (e) are DF scattering spectra of coupled hybrid systems at the first three rows. The dark dashed lines and gray dashed lines represent the absorption peak positions of A and B excitons, respectively. Red dashed curves trace (guide to the eye) the dispersion of plexciton branches.}% the caption of figure

%\label{f1}%the citation of the figure in the manuscript
\end{figure}

%\textbf{Strong coupling in MoS$_2$ layers}.
To verify the coupled system is in strong coupling regime, tuning the plasmon mode to cross the energy of A exciton is required. In most cases, the tuning comes from randomly distributed nanoparticles with different sizes \cite{chikkaraddy2016single,qin2020revealing,stuhrenberg2018strong,nanocube}. Since the energy of plasmon is sensitive to the particle size, this inevitably limits the reliability and repeatability of the system. Particularly for MoS$_2$, the splitting of spin-orbit coupling is close to the value of Rabi splitting, complicating the study of such system because of the emission of B exciton. Here, increasing the gap distances of bowtie resonators has been used to tune gap modes in a more moderate way than tuning sizes (Fig. S10 and Fig. S11 in the SI).

Figure~\ref{p2}a shows a bright-field image of the bowties covered with large area of MoS$_2$ layers. It can be seen that the whole area is flat without any wrinkles due to the above exfoliation and transfer techniques. The smallest gap distances of bowties are $20\pm2$ nm and increase by approximate 5 nm for each step from left to right (Fig. S1 and Fig. S2 in the SI). The red dashed box indicates that there are five rows of bowties covered well by flakes, labelled with Roman numerals from one to five. Figure~\ref{p2}b shows the dark-field image of the first three rows. The clear and bright spots of hybrid nanostructures indicate the TMDs layers are combined well with the plasmonic resonators without extra wrinkles and bubbles.

\begin{figure}
\centering
\includegraphics[scale=0.5]{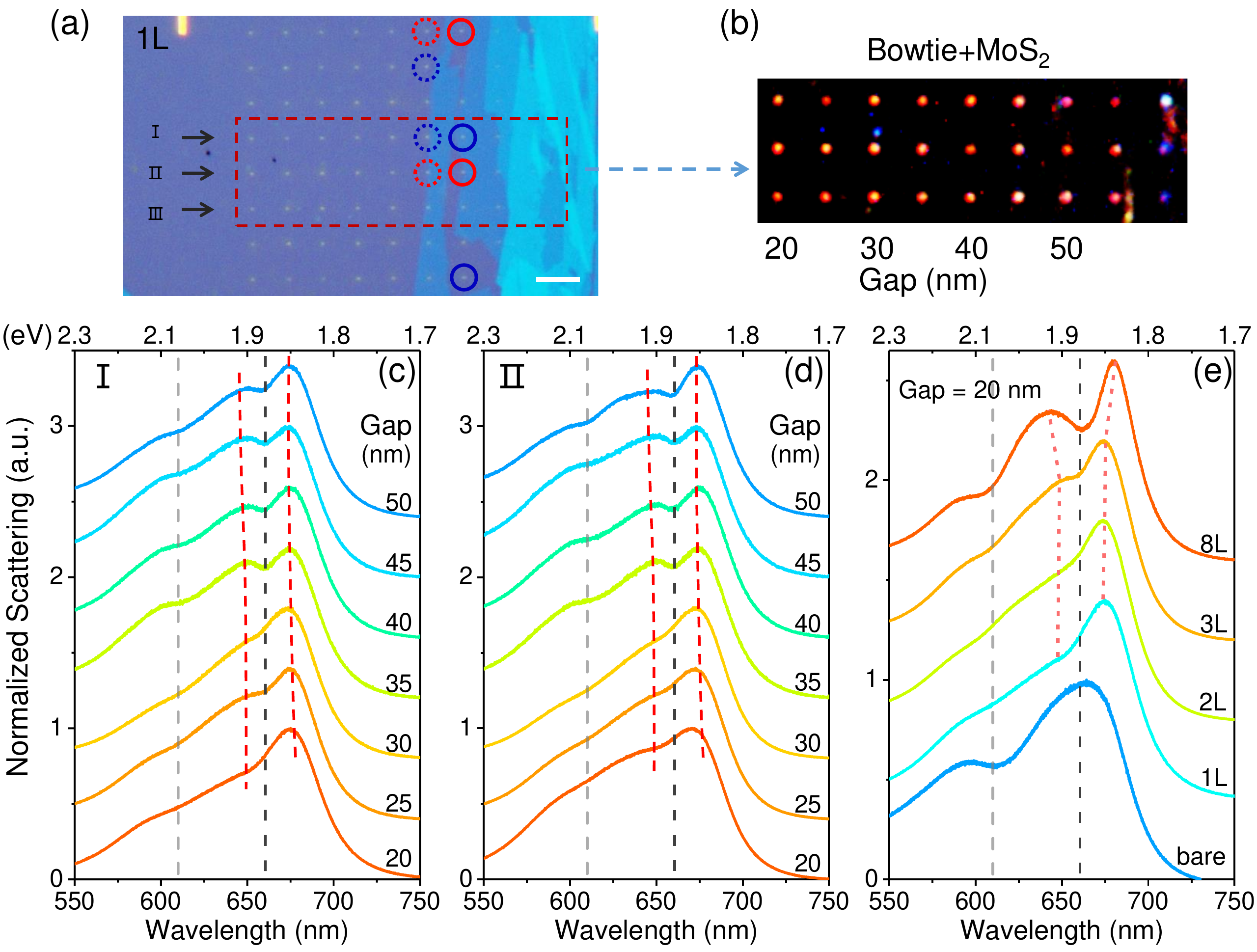} %scale is the size of the figure, you can modify it. f1_v1 is the name of figure.eps
\caption{\label{p3}Exciton-plasmon coupling in different layers of MoS$_2$. (a) Bright-field image of monolayer MoS$_2$ on bowties. The parameters of bowties from left to right are the same with that above. (b) The corresponding dark-field image in (a) (scale bar: 4 ${\mu}m$). (c) and (d) are DF scattering spectra of line \uppercase\expandafter{\romannumeral1} and \uppercase\expandafter{\romannumeral2}. The spectra from devices marked with circles in \uppercase\expandafter{\romannumeral1} and \uppercase\expandafter{\romannumeral2} are replaced from other devices with the same circles covering with monolayer. The dark dashed lines and gray dashed lines represent the absorption peak positions of A and B excitons, respectively.  Red dashed curves (guide to the eye)  trace the dispersion of plexciton branches. (e) Scattering spectra of different layers of MoS$_2$ on bowties with the same gap distance about 20 nm.}% the caption of figure
%\label{f1}%the citation of the figure in the manuscript
\end{figure}

The normalized scattering spectra from rows \uppercase\expandafter{\romannumeral1}, \uppercase\expandafter{\romannumeral2}, \uppercase\expandafter{\romannumeral3} (Fig.~\ref{p2}c, d, e) and \uppercase\expandafter{\romannumeral4}, \uppercase\expandafter{\romannumeral5} (Fig. S15 in the SI) all show similar behaviors, indicating a good robustness. When the gap distance is around 20 nm, an obvious double-peaked splitting around the position of A exciton is observed, representing the energy of the upper plexciton branches (UPB) and lower plexciton branches (LPB). It is worth noting that the transferred MoS$_2$ layers will change the dielectric environment of resonators and result in a slight redshift of plasmon mode due to the dielectric screening effect \cite{zheng2017manipulating, xu2003wavelength}. With the increase of gap distance, the longitudinal mode continuously blueshifts and eventually crosses the excitonic transition (see Sec. II of the SI for the discussion of energy tuning between exciton and plasmon). To extract the peak energies of UPB and LPB with Lorentzian fitting method, we fix the resonance of B exciton at about the 2.0 eV according to the reflection spectrum in Figure 1d (see Fig. S17 in the SI for fitting details). The red dashed curves in the scattering spectra trace the dispersion of plexciton branches, showing that UPB is getting closer to the resonance of B exciton as the gap increases but doesn't overlap with it, which means the longitudinal plasmon mode only couples with A exciton of MoS$_2$  unambiguously here.

The similar splitting properties are also observed in the devices covered with monolayer. Figure~\ref{p3}a shows a bright-field image of the bowtie resonators with a large area of monolayer flake. The corresponding dark-field image is shown in Fig.~\ref{p3}b, where the resonator parameters are the same as those above. Figure~\ref{p3}c and d show the measured scattering spectra of line \uppercase\expandafter{\romannumeral1} and \uppercase\expandafter{\romannumeral2} with consistent spectral features (see Fig. S16 in the SI for line \uppercase\expandafter{\romannumeral3}). It can be seen that  a relatively small splitting is observed compared with that of 8 layers. The splitting depending on layer thickness is also studied. As shown in Fig.~\ref{p3}e, the magnitude of splitting in spectra increases with the layer thickness with a gap distance at $\sim$ 20 nm, implying an increase of coupling strength with the number of layers. Although the strong coupling for all different layers has not been experimentally achieved, the splitting difference for different layer is obvious.

\begin{figure}
\centering
\includegraphics[scale=0.4]{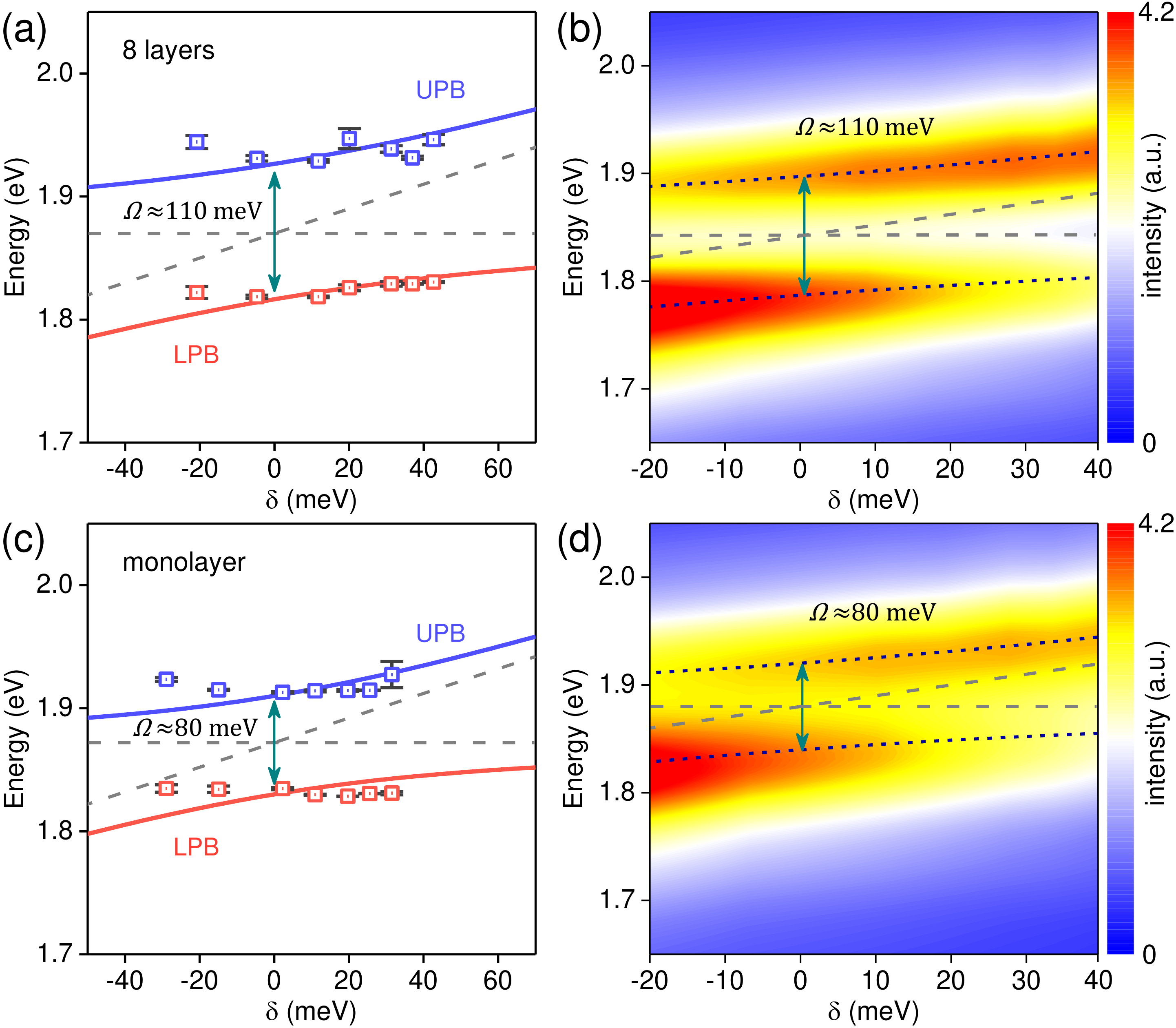} %scale is the size of the figure, you can modify it. f1_v1 is the name of figure.eps
\caption{\label{p4}Dispersion of plexciton and the corresponding calculation results. (a) (c) The average energies of the UPB (blue squares) and LPB (magenta squares) of the several sets of data as a function of the detuning for the coupling with 8 layers and monolayer respectively. The curves in both (a) and (c) are fitted to JC model, giving Rabi splittings of 110 meV and 80 meV, respectively. (b) (d) Calculated cross section of individual bowtie resonators covered by 8-layer and monolayer MoS$_2$ as a function of the detuning. Exciton energy and plasmon energy are denoted by gray dashed lines.}
\end{figure}

The coupled system can be described by the simplified Jaynes-Cummings model (JC model) given by \cite{torma2014strong,qin2020revealing,qian2018two}:
\begin{equation}
\omega_{\pm} = \frac{1}{2}(\omega_{pl}+\omega_{ex})\pm\sqrt{g_c^2+\frac{1}{4}\delta^2}
\end{equation}
where $\omega_{pl}$, $\omega_{ex}$ are the energies of plasmon and exciton respectively, $\delta = \omega_{pl}-\omega_{ex}$ is the detuning, and $g_c$ represents the coupling strength. A fit to the UPB and LPB peak energies using JC model is shown in Fig.~\ref{p4}a and c. As we can see, the errors of peak energies between different group of spectra are very small, showing the high robustness and reproducibility of the coupled system. The JC model fits to the peak energies show a Rabi splitting ($\Omega = 2g_c|_{\delta=0}$) about 110 meV for 8-layer devices, which satisfies the criteria for strong coupling ($\Omega > (\Gamma_{pl}+\Gamma_{ex})/2$). While the Rabi splitting for the monolayer is approximately  $80$ meV, indicating the system is in intermediate-coupling regime ($\Omega > (\Gamma_{pl}-\Gamma_{ex})/2$) \cite{leng2018strong}. Numerical calculations provide another piece of evidence for our observations. By modeling the excitonic dielectric permittivity of the MoS$_2$ as a Lorentz oscillator, we calculated the scattering cross section of hybrid structures with changing the gap distance from 20 to 50 nm (Fig.~\ref{p4}b and d), showing an anticrossing of two normal modes.

Comparing with the calculated results, the intensity of UPB in experiment seems to be always lower than that of LPB, which can be due to the rapid attenuation of the gap mode with the increase of the gap and the non-negligible emission of uncoupled A exciton outside the nanocavity. The JC model fits and calculations also reveal the moderate anticrossing behavior of UPB and LPB in Fig.~\ref{p4}a and c. Due to the small tuning range of plasmon mode with increasing gap sizes and the strong coupling of exciton and plasmon optimized to near resonance, the energies of UPB and LPB move more slowly compared with that of plasmon mode during tuning process, making the data points looks horizontal and less curved. Additionally, the small number of excitons contributing to the coupling with gap mode also accounts for the mismatch between JC model fits and plexciton branches because of the strong background signal. For example, the emission of massive uncoupled excitons near bowtie in scattering spectra will affect the extraction of plexciton branches.

To evaluate the exciton number evolved in the strong couling, we use $g_c=\sqrt{N}\bm{\mu}\cdot\bm{ E_{vac}}=\sqrt N \bm{\mu}\cdot|E_{vac}|\bm{K}$, where $N$ is the effective exciton number coherently contributing to the interaction with the cavity, $\bm{\mu}$ is exciton transition dipole moment, $|E_{vac}|=\sqrt{\hbar\omega/2\varepsilon_r\varepsilon_0 V_m}$ is the vacuum field amplitude \cite{zengin2015realizing,torma2014strong} and $|\bm{K}|$ is unit vector, satisfy $|\bm{K}|=1$. Because the exciton dipole strength in TMDs is highly anisotropic \cite{liang1973,schuller2013orientation} and has an out-of-plane component in multilayers, the coupling strength can be written as
\begin{equation}
g_c= \sqrt N |E_{vac}|( \bm{\mu_{xy}}+ \bm{\mu_{z}})\cdot(\bm{K_{xy}}+\bm{K_{z}})
\end{equation}
where $\bm{\mu_{xy}}$ and $\bm{\mu_{z}}$ represent the in-plane and out-of-plane dipole moments respectively, and $\bm{K}=\bm{K_{xy}}+\bm{K_{z}}$ with $\bm{K_{xy}}$ being parallel to the two-dimensional semiconductors plane and $\bm{K_{z}}$ being perpendicular to the plane. Therefore, coupling strength can be expressed as a form of contribution from in-plane and out-of-plane dipole moments: $g_c= \sqrt N |E_{vac}| (\mu_{xy}K_{xy}+ \mu_{z}K_{z})$. Here, we define the ratio of in-plane field as $\beta_{xy}=|K_{xy}|^2$, which represents the ratio of an integral of in-plane field components to the total electric field and can be numerically calculated by:
\begin{equation}
\beta_{xy}=\int_{V_e} {\frac {E_x^2 + E_y ^2}{E_x^2 + E_y ^2+E_z^2}} dV_e
 \end{equation}
 where $V_e$ represents the volume of excitonic material. Similarly, we can get the ratio of out-of-plane field $\beta_{z}$ with $\beta_{xy}+\beta_{z}=1$. Finally, the coupling strength can be written as
\begin{equation}
g_{c} = \sqrt{\frac{N\hbar\omega}{2\varepsilon_r\varepsilon_0 V_m}}(\mu_{xy}\sqrt{\beta_{xy}}+\mu_{z}\sqrt{\beta_{z}})
\end{equation}

\begin{figure}
\centering
\includegraphics[scale=0.5]{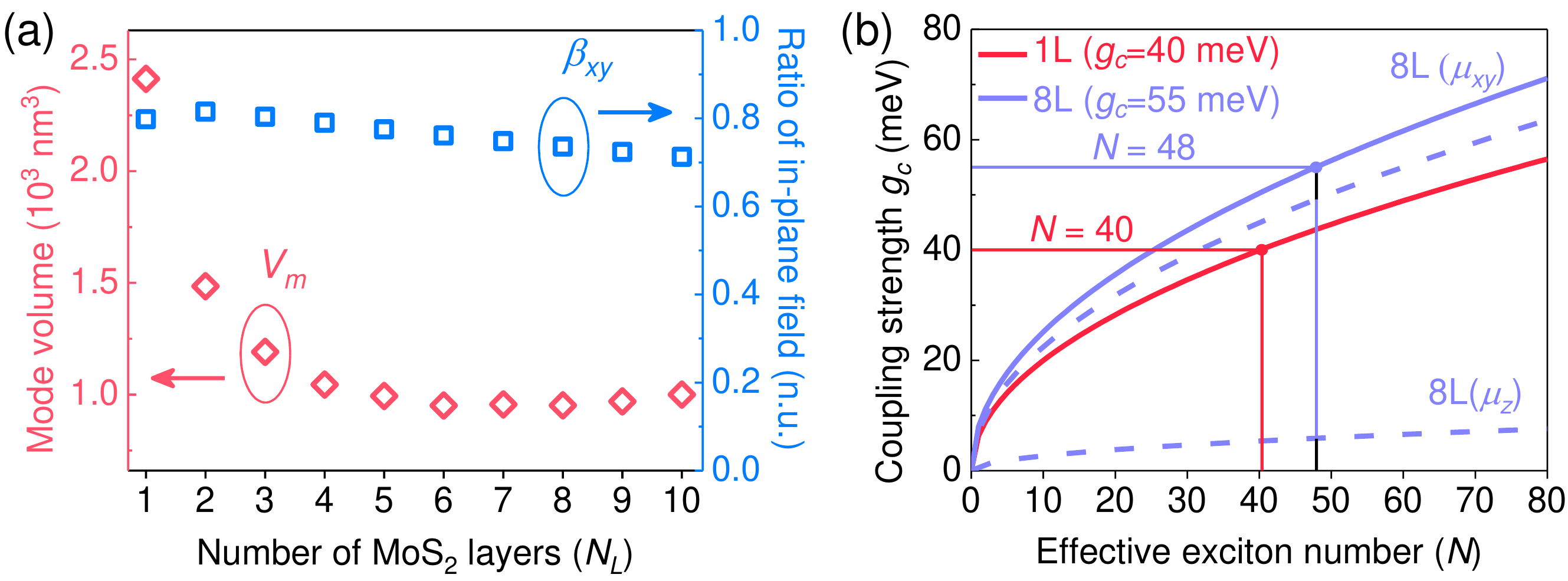} %scale is the size of the figure, you can modify it. f1_v1 is the name of figure.eps
\caption{\label{p5}The effective exciton number of the coupled system. (a) Mode volume $V_m$ and ratio of in-plane field $\beta_{xy}$ as a function of the number of MoS$_2$ layers with the same gap distance about 20 nm. (b) Coupling strength $g_c$ as a function of the effective exciton number ($N$) for monolayer and 8-layer systems. The dashed blue lines show the coupling strength components of in-plane ($\mu_{xy}$) and out of the plane  ($\mu_z$).}% the caption of figure
%\label{f1}%the citation of the figure in the manuscript
\end{figure}

 Figure~\ref{p5}a shows the mode volumes as a function of the layer number. The mode volume of bowtie nanocavity is about $10^3 \ nm^3$ for monolayer, which is comparable to the current optimal results for a single nanoparticle such as gold bi-pyramids \cite{stuhrenberg2018strong} or utrasmall gold nanorod \cite{SinglePlasmonicNanorod}, and even smaller in the case with multiple layer covering. With layer number increasing, the mode volume gradually decreases from 2413 to 951 $nm^3$ and then saturates. This means that the electric field is strengthened and more tightly confined in TMDs layers, which explains why larger splittings are observed in multilayers. The ratio of an integral of  in-plane field to total field is about $71\sim80\%$ in different layers, as shown in Fig.~\ref{p5}a, confirming that the dominant electric field component of gap plasmon in our system is the in-plane component $E_{xy}$.

%The transition dipole moments of different MoS$_2$ layers are obtained from the absorbance measurements (see Sec. III of the SI for more details), indicating $\mu \approx$ 7.36 Debye ($D$) for monolayer MoS$_2$. For the case of several layers, we determined the in-plane dipole moment using $\mu_{xy} \propto \sqrt{abs}$, where $abs$ is spectrally integrated for the A excitonic transition \cite{wang2019limits}. Using the value of monolayer, we obtain the in-plane dipole moment $\mu_{xy}^{8l} \approx$ 5.07 $D$ and out of plane $\mu_{z}^{8l} \approx$ 1.01 $D$.

In order to estimate the number of excitons involved in coupling, transition dipole moment of the excitons in TMDs layer is another significant parameter. Here, we adopt two methods i.e. quantum well method and absorbance measurements to estimate this value. The quantum well method regards the 2D TMDs layer as quantum-well structures similar to III-V semiconductors \cite{asada1984gain} and takes into account that the electron of TMDs has a large effective mass $m_c$ around the K point, giving $\mu =\frac{e\hbar}{2E_0}[\frac{E_g(E_g+\Delta_0)}{E_g+2\Delta_0/3}(\frac{1}{m_c}-\frac{1}{m_0})]^{\frac{1}{2}}$, where $E_0$ is the transition energy of the exciton, $E_g$ is the band gap and $\Delta_0$ is the spin-orbit splitting in the valence band. The absorbance measurements consider the relationship between the 2D susceptibility of excitons with 1$s$ hydrogen-like wave function and absorption $A_{ex}^{2D}$ of the thin layer \cite{zheng2017manipulating,vasilevskiy2015exciton}, giving $A_{ex}^{2D} =\frac{4 \eta_0\omega_{ex}}{\pi a_B^2\Gamma_{ex}}\mu^2$, where $\eta_0$ is the free space impedance, $a_B$ is the Bohr radius of exciton, $\omega_{ex}$ and $\Gamma_{ex}$ are the energy and linewith of exciton, respectively. Both methods give similar dipole moment values for TMDs layers, such as 7.53 Debye ($D$) and 5.63 $D$ for monolayer WS$_2$ and 7.51 $D$ and 7.36 $D$ for monolayer MoS$_2$, respectively (see Sec. III of the SI for more details). It should be noted that our calculation result is much smaller than the value of 56 $D$ reported in the literature \cite{0Valley}, where the reduced Plank constant ($\hbar$) should have been used as discussed in the SI. For the case of several layers, we determined the in-plane dipole moment using $\mu_{xy} \propto \sqrt{abs}$, where $abs$ is spectrally integrated for the A excitonic transition \cite{wang2019limits}. Using the value of monolayer, we obtain the in-plane dipole moment $\mu_{xy}^{8l} \approx$ 5.07 $D$ and out of plane $\mu_{z}^{8l} \approx$ 1.01 $D$.

\begin{table}[h]
\caption{Reported effective exciton numbers in different plasmonic cavities.}
\small
\label{ts1}
\begin{tabular}{p{5cm}p{2cm}p{1.5cm}p{2.5cm}p{2.5cm}p{2.5cm}}\\ % four columns, alignment for each
	\hline
	Structure  & Materials & $\Omega$ (meV) & N ($\mu_{0}$\textsuperscript{*}) & N ($\mu_{qw}$\textsuperscript{*})
&N ($\mu_{ab}$\textsuperscript{*})\\
	\hline

Single gold nanoprism on gold film (gap plasmon)\cite{qin2020revealing} & WS$_2$ &  76& 2 \newline (56 $D$)&111 \newline(7.53 $D$)&198 \newline(5.63 $D$)\\ \\

Single gold dimer (gap plasmon)\cite{liu2021plasmon} & WS$_2$ & 115.2-128.6 & 4.67-7.69 \newline(56 $D$) &258-425 \newline(7.53 $D$)&462-761 \newline(5.63 $D$) \\ \\

Single silver nanocube on silver film (gap plasmon) \cite{nanocube} & WS$_2$ & 145 & 130 \newline(56 $D$) &7190 \newline(7.53 $D$)&12862 \newline(5.63 $D$) \\ \\

Silver nanoparticle array \cite{wang2019limits} & WS$_2$ & 52 & 3000 \newline(50 $D$) & $\sim$ 132000 \newline(7.53 $D$)& $\sim$236000\newline (5.63 $D$) \\ \\

Single silver nanorod \cite{zheng2017manipulating} & WSe$_2$ & 49.5 & 4100 \newline(7.67 $D$) \\ \\

Single gold nanorod \cite{SinglePlasmonicNanorod}& WS$_2$ & 106 & $\sim$12  &225 \newline(7.53 $D$)& 403 \newline(5.63 $D$) \\ \\

Our work: Single gold bowtie (gap plasmon) & MoS$_2$ & 80-110 & & 38 \newline (7.51 $D$)& 40\newline (7.36 $D$)\newline 48 \newline(8 layer)\\ \\
	\hline
\end{tabular}
\textsuperscript{*} $\mu_{0}$ is the transition dipole moment in refs, $\mu_{qw}$ is corrected transition dipole moment with quantum well model and $\mu_{ab}$ is calculated with absorption spectra.
\end{table}

Fig.~\ref{p5}b shows the calculated coupling strength $g_c$ as a function of the effective exciton number $N$ for monolayer and 8 layers at resonance. We found the effective exciton number is compressed down to $N\sim40$ for the case of monolayer and $N\sim48$ in multilayers, indicating a small exciton number in such coupling system with plasmon modes and excitons in two-dimensional semiconductors. Table 1 shows the comparison of the coupled systems with small exciton number with some previous reports, in which the exciton numbers with corrected dipole moments are also included (see Sec. III of the SI for more details). The effective exciton numbers are much larger with corrected dipole moment than those as reported \cite{qin2020revealing,nanocube,wang2019limits,liu2021plasmon}. The numbers involved in single nanorod structure\cite{SinglePlasmonicNanorod} are recalculated by the formula $g_c=\sqrt{N}\bm{\mu}\cdot\bm{ E_{vac}}$ at zero tuning. The calculated exciton numbers are also much larger than that as claimed. The small number of exciton in our experiment also explains the mismatch between JC model fitting and experimental results in Fig.~\ref{p4}a and c due to the influence of emission of massive uncoupled excitons in layers around the bowtie. For multilayers, the contribution to coupling strength is only $12\%$ ($\mu_{z}\sqrt{\beta_{z}}/\mu_{xy}\sqrt{\beta_{xy}}$) of the in-plane component as shown in Fig.~\ref{p5}b, indicating a selective coupling between the larger in-plane exciton dipole moment and the dominate in-plane field of gap mode in our configuration as designed. Furthermore, the effective number in our system can be reduced more by further reducing the gap distance of bowtie until it is comparable to the exciton coherence length ($\sim4$ nm). When the gap size is smaller than the exciton coherence length, the out-of-plane component of gap mode located in the two tips of bowtie will drive the exciton dipole with opposite phase as shown in Fig. \ref{p1}c, which prevents the further coupling enhancement.

\section{\label{sec4}Conclusion:}

In summary, we have demonstrated a plasmon-exciton strong coupling between individual bowtie resonators and MoS$_2$ layers, with the effective exciton number contributing to the coupling down to 40 in monolayer and 48 in few layers. Such a small exciton number in the plexciton system shows an opportunity to study the interaction between cavity and many emitters, and to achieve potentially a strong coupling between single exciton and plasmon in two-dimensional materials with a small mode volume \cite{PhysRevLett.126.257401}. Moreover, we also demonstrate an universal method to obtain robust and reproducible plasmon-exciton interaction by utilizing a gold-assisted mechanical exfoliation method and wet transfer techniques, which paves a way to integrate the plexciton system into photonic devices and exploit novel quantum and nonlinear optic effects at room temperature.

\section{Supporting Information}
\subsection{Supporting Information I. Sample Fabrication and characterizations}
\subsection{Supporting Information II. Simulation and optimization of plasmon mode in nanocavity}
\subsection{Supporting Information III. Calculation of effective exciton number}
\subsection{Supporting Information IV. Extra data of DF Scattering Spectra of coupled structures with monolayer and 8-layer MoS$_2$}

\section{acknowledgement}
This work was supported by the National Key Research and Development Program of China (Grant No. 2021YFA1400700, No. 2019YFA0308000 and No. 2018YFA0704201), the National Natural Science Foundation of China (Grants No. 62025507, No. 11934019, No. 11721404, No. 11874419, No. 11874405, No. 62022089, No.62175254 and No. 12174437), the Key-Area Research and Development Program of Guangdong Province (Grant No. 2018B030329001), the Strategic Priority Research Program (Grant No. XDB28000000 and No. XDB33000000), the Youth Innovation Promotion Association of CAS (2019007) of the Chinese Academy of Sciences. We thank Professor Masahiro Asada for helpful discussions.

\end{document}